\documentclass{article}
\usepackage[utf8]{inputenc}
\usepackage{mathrsfs}
\usepackage{fullpage}
\usepackage{amsfonts}
\usepackage{graphicx}
\usepackage{amsmath}
\usepackage{amssymb}
\usepackage{float}
\usepackage{subfigure}
\usepackage{rotating}
\usepackage{xcolor}
\usepackage{natbib}
\usepackage{epstopdf}
\usepackage{enumerate}
\usepackage{url}
\usepackage[space]{grffile}
\usepackage{latexsym}
\usepackage{textcomp}
\usepackage{longtable}
\usepackage{graphicx} 
\usepackage{tabulary}
\usepackage{booktabs,array,multirow}
\usepackage{epstopdf}
\usepackage{epsfig,fancyheadings}
\usepackage{sectsty, secdot}
\usepackage{amsthm}
\usepackage{bm,bbm}
\usepackage{setspace}
\usepackage{geometry,algorithm,algpseudocode}
\usepackage[colorlinks,citecolor=blue,urlcolor=blue,linkcolor = blue]{hyperref}

\newtheorem{theorem}{Theorem}

\def\E{\mathrm{E}}
\def\Pr{\mathrm{P}}
\def\var{\mathrm{var}}

\def\cov{\mathrm{cov}}

\def\cd{\mathop{\rightarrow}\limits^{d}}
\def\mR{\mathbb{R}}

\def\H{{\bf H}}
\def\I{{\bf I}}

\def\P{{\bf P}}

\def\X{{\bf X}}
\def\Y{{\bf Y}}

\def\e{{\bm e}}

\def\u{{\bf u}}

\def\bms{{\bf \Sigma}}

\allowdisplaybreaks[4]

\title{Adaptive Test Procedure for High Dimensional Regression Coefficient}
\author{Ping Zhao\\
School of Mathematical Sciences, Tiangong University\\
Fengyi Song and Huifang Ma\\
School of Statistics and Data Science, Nankai University}
\date{\today}

\begin{document}

\maketitle

\begin{abstract}
We develop a unified $L$-statistic testing framework for high-dimensional regression coefficients that adapts to unknown sparsity.
The proposed statistics rank coordinate-wise evidence measures and aggregate the top $k$ signals, bridging classical max-type and sum-type tests. We establish joint weak convergence of the extreme-value component and standardized $L$-statistics under mild conditions, yielding an asymptotic independence that justifies combining multiple $k$'s.
An adaptive omnibus test is constructed via a Cauchy combination over a dyadic grid of $k$, and a wild bootstrap calibration is provided with theoretical guarantees. Simulations demonstrate accurate size and strong power across sparse and dense alternatives, including non-Gaussian designs.

\end{abstract}

\section{Introduction}\label{sec:introduction}

Modern applications routinely generate regression problems where the number of covariates is comparable to, or far exceeds, the sample size. In such regimes, a central inferential task is to test whether a large collection of regression coefficients is jointly null, possibly after adjusting for a small set of nuisance covariates. This type of global or blockwise testing problem arises broadly in genomics, neuroimaging, and finance, where scientific questions are often naturally formulated as whether a high-dimensional set of predictors contributes explanatory power beyond a baseline model.

A key difficulty is that the power-optimal testing strategy depends strongly on the  {sparsity pattern} of the alternative. When signals are sparse (a few coordinates carry most of the effect), max-type procedures that target the largest marginal evidence are typically effective; see, for example, the max-type methodology and related asymptotic independence phenomena developed in \citet{feng2024asymptotic}. In contrast, when signals are dense (many weak effects), sum/energy-type procedures are often preferred, including the global testing framework of \citet{GoemanVanDeGeerVanHouwelingen2006JRSSB}, its extensions and type-I error control in generalized linear models \citet{GoemanVanHouwelingenFinos2011Biometrika}, and related sum-type tests for high-dimensional regression coefficients such as \citet{ZhongChen2011JASA} and \citet{LanWangTsai2014AISM}. In practice, however, the sparsity level is rarely known a priori, and procedures tuned to a single regime may suffer substantial power loss when the underlying alternative deviates from that regime. \cite{cui2018test} used refitted cross-validation variance estimation to avoid the overestimation of the variance and enhance the empirical power. \cite{guo2022conditional} proposed a two-stage conditional U-statistic test with screening procedure based on a random data-splitting strategy to enhance the empirical power.

This motivates adaptive testing strategies that can maintain sensitivity across a wide spectrum of alternatives. A natural device is to construct a  {family} of statistics indexed by a tuning parameter that interpolates between sparse- and dense-oriented behavior, and then combine information across this index set. In this paper, we pursue this idea via an $L$-statistic framework: we rank coordinate-wise evidence measures and aggregate the largest $k$ components, thereby forming a continuum of top-$k$ $L$-statistics that smoothly bridge max-type and sum-type tests. This perspective is closely related to recent advances on adaptive $L$-statistics for high-dimensional testing developed in \citet{ma2024adaptive}, and it allows us to formulate a single methodology that can adapt to unknown sparsity in a principled manner.

A main theoretical challenge for multi-$k$ aggregation is the substantial dependence among $\{L_k\}$ across nearby values of $k$, as well as the interaction between the extreme-value behavior of the top order statistics and the Gaussian-type fluctuations that arise when $k$ diverges. Without a careful joint characterization under the null, naive combination of multiple $k$’s may lead to conservative behavior or size inflation. Building on the asymptotic independence insights in \citet{feng2024asymptotic} and the expected-shortfall-based normal approximation theory in \citet{ma2024adaptive}, we establish joint weak convergence results that clarify how the extreme-value component and the standardized $L$-statistics factorize under the null. This asymptotic independence provides a rigorous foundation for combining evidence across different $k$ values while retaining valid type-I error control.

Methodologically, we propose an omnibus adaptive test by aggregating $p$-values corresponding to a dyadic grid of truncation levels $k$, supplemented by a small number of extreme-type components. The resulting procedure is computationally simple—since all component $p$-values can be obtained within a unified bootstrap routine—yet it is designed to be robust across sparsity regimes: smaller $k$ emphasize rare/strong signals, whereas larger $k$ accumulate widespread/weak signals. To improve finite-sample calibration and to accommodate non-Gaussian designs, we employ a wild bootstrap scheme for approximating the null distributions of the entire $L$-statistic family. Although part of our limit theory leverages Gaussian structure for technical convenience, our simulations indicate that the proposed method continues to perform well under a range of non-Gaussian distributions.

The remainder of the paper is organized as follows. Section~2 introduces the family of top-$k$ $L$-statistics and the bootstrap calibration. Then, we develops the asymptotic theory in both fixed-$k$ and diverging-$k$ regimes and establishes the joint limit results that motivate our adaptive combination. Section~3 reports simulation studies comparing the proposed procedure with representative max-type and sum-type competitors. Section~4 concludes with a discussion and possible extensions.

\section{Test Procedures}

Let \(\boldsymbol{X}=(\boldsymbol{X}_1,\ldots,\boldsymbol{X}_n)^\top\) be i.i.d.\ \(p\)-dimensional covariates and \(\boldsymbol{Y}=(Y_1,\ldots,Y_n)^\top\) be the corresponding independent responses. For notational convenience, assume \(E(\boldsymbol{X}_i)=\mathbf 0\). We partition each covariate vector as
\[
\boldsymbol{X}_i=\big(\boldsymbol{X}_{ia}^\top,\boldsymbol{X}_{ib}^\top\big)^\top,\qquad 
\boldsymbol{X}_{ia}\in\mathbb R^{q},\ \boldsymbol{X}_{ib}\in\mathbb R^{p-q},
\]
where \(q<n\) is the dimension of nuisance covariates and \(p\) may be much larger than \(n\). Consider the linear regression model
\begin{align}\label{model}
Y_i=\boldsymbol{X}_i^\top\boldsymbol{\beta}+\varepsilon_i
=\boldsymbol{X}_{ia}^\top\boldsymbol{\beta}_a+\boldsymbol{X}_{ib}^\top\boldsymbol{\beta}_b+\varepsilon_i,
\end{align}
where \(\boldsymbol{\beta}=(\boldsymbol{\beta}_a^\top,\boldsymbol{\beta}_b^\top)^\top\in\mathbb R^{p}\), with \(\boldsymbol{\beta}_a\in\mathbb R^{q}\) and \(\boldsymbol{\beta}_b\in\mathbb R^{p-q}\). The noises \(\{\varepsilon_i\}_{i=1}^n\) are independent with \(E(\varepsilon_i)=0\), \(\mathrm{Var}(\varepsilon_i)=\sigma^2\), and are independent of \(\boldsymbol{X}\).

Our goal is to test whether the high-dimensional block \(\boldsymbol{\beta}_b\) is zero:
\begin{align}\label{hy}
H_0:\boldsymbol{\beta}_b=\mathbf 0 \quad \text{vs.}\quad H_1:\boldsymbol{\beta}_b\neq \mathbf 0,
\end{align}
which reduces to the global test \(H_0:\boldsymbol{\beta}=\mathbf 0\) when \(\boldsymbol{X}_{ia}\equiv \mathbf 0\).

To remove the nuisance component \(\boldsymbol{\beta}_a\), define the design matrices
\[
\X_a=(\boldsymbol{X}_{1a},\ldots,\boldsymbol{X}_{na})^\top,\qquad 
\X_b=(\boldsymbol{X}_{1b},\ldots,\boldsymbol{X}_{nb})^\top,\qquad m=p-q,
\]
and the projection matrix onto the column space of \(\X_a\),
$\H_a=\X_a(\X_a^\top\X_a)^{-1}\X_a^\top.$
We then residualize \(\X_b\) with respect to \(\X_a\) by
$\tilde \X_b=(\I_n-\H_a)\X_b=\big(\tilde\X_{\cdot 1},\ldots,\tilde\X_{\cdot m}\big),$
and define the corresponding residuals and variance estimator
\[
\hat{\bm \varepsilon}=(\I_n-\H_a)\Y,\qquad 
\hat\sigma^2=(n-q)^{-1}\hat{\bm \varepsilon}^\top\hat{\bm \varepsilon}.
\]

Based on the residualized design, we consider the collection of standardized score-type statistics
\[
Z_j=\frac{\tilde\X_{\cdot j}^\top\hat{\bm \varepsilon}}{\hat\sigma\|\tilde\X_{\cdot j}\|_2},
\qquad W_j:=Z_j^2,\qquad j=1,\ldots,m,
\]
and let \(W_{(1)}\ge \cdots\ge W_{(m)}\) be the order statistics. For \(1\le k\le m\), define the top-\(k\) \(L\)-statistic
\[
L_k=\sum_{j=1}^k W_{(j)}.
\]
This family provides a unified bridge between two classical regimes: when \(k=1\), \(L_k\) reduces to an \(\ell_\infty\)-type (maximum) statistic, closely related to the max-type test in \cite{feng2024asymptotic}; when \(k=m\), \(L_k\) becomes an \(\ell_2\)-type (sum/energy) statistic, connecting to sum-type procedures such as \cite{GoemanVanDeGeerVanHouwelingen2006JRSSB,GoemanVanHouwelingenFinos2011Biometrika,ZhongChen2011JASA,LanWangTsai2014AISM}. More generally, intermediate choices of \(k\) interpolate between sparse and dense alternatives, thereby motivating the study of both fixed-\(k\) and diverging-\(k\) asymptotics.

In what follows, we derive the null limiting distributions of \(L_k\) in two complementary scenarios: (i) \(k\) is fixed, leading to an extreme-value/Poisson point-process limit for the top order statistics; and (ii) \(k\) diverges with \(m\), for which \(L_k\) admits an asymptotic normal approximation. These results form the theoretical basis for constructing adaptive procedures that combine evidence across multiple \(k\)’s.

\subsection{Fixed parameter $k$}
Define the partial covariance matrix
$
\boldsymbol{\Sigma}_{b\mid a}
:=E\!\left[\mathrm{Cov}\!\left(\boldsymbol{X}_{ib}\mid \boldsymbol{X}_{ia}\right)\right]
=\big(\sigma^{*}_{jk}\big)\in\mathbb R^{m\times m}.
$
Without loss of generality, we normalize \(\boldsymbol{\Sigma}_{b\mid a}\) so that \(\sigma^{*}_{jj}=1\) for all \(j\in\{1,\ldots,m\}\).
Let \(\mathbf{B}\) be the population regression coefficient of \(\boldsymbol{X}_{ib}\) on \(\boldsymbol{X}_{ia}\),
$
\mathbf{B}:=\mathrm{Cov}(\boldsymbol{X}_{ib},\boldsymbol{X}_{ia})\,
\big[\mathrm{Cov}(\boldsymbol{X}_{ia})\big]^{-1}\in\mathbb R^{m\times q},
$
and define the residual (innovation) vector
$
\boldsymbol{X}_{ib}^*:=\boldsymbol{X}_{ib}-\mathbf{B}\boldsymbol{X}_{ia}=(X_{i1}^*,\cdots,X_{im}^*)^\top\in\mathbb R^{m}.
$
Since \(E(\boldsymbol{X}_i)=\mathbf 0\), it follows that \(E(\boldsymbol{X}_{ib}^*)=\mathbf 0\) and
$
\mathrm{Cov}(\boldsymbol{X}_{ib}^*)=\boldsymbol{\Sigma}_{b\mid a}.
$
Finally, collect the residuals into the matrix
$
\mathbf{X}_b^*:=\big(\boldsymbol{X}_{1b}^*,\ldots,\boldsymbol{X}_{nb}^*\big)^\top\in\mathbb R^{n\times m}.
$

For a set of multivariate random vectors $\boldsymbol{z}=\left\{z_j: j \geq 1\right\}$ and integers $a<b$, let $\mathcal{Z}_a^b$ be the $\sigma$-algebra generated by $\left\{z_j: j \in[a, b]\right\}$. For each $s \geq 1$, define the $\alpha$-mixing coefficient $\alpha_Z(s)=\sup _{t \geq 1}\left\{|P(A \cap B)-P(A) P(B)|: A \in \mathcal{Z}_1^t, B \in \mathcal{Z}_{t+s}^{\infty}\right\}$.

 The conditions we will use later on are stated below.
\begin{itemize}
\item[(A1)] $\boldsymbol{X}_{i b}^* \sim N\left(\mathbf{0}, \boldsymbol{\Sigma}_{b \mid a}\right)$ and the diagonal entries of $\boldsymbol{\Sigma}_{b \mid a}$ are all equal to $1 ;$

\item[(A2)]There exists a constant $\tau>1$ such that $\tau^{-1}<\lambda_{\text {min }}\left(\boldsymbol{\Sigma}_{b \mid a}\right) \leqslant \lambda_{\text {max }}\left(\boldsymbol{\Sigma}_{b \mid a}\right)<\tau$. There exists $r_1>0$ such that $\max_{1\le i,j\le m} \sigma_{ij}^* \le r_1\le 1$.

\item[(A3)] Suppose $p=o\left(n^3\right), q=o(p), q \leqslant n^\delta$ for some $\delta \in(0,1)$ and $E\left(\left|\varepsilon_1\right|^{\ell}\right)<\infty$ with $\ell=14(1-\delta)^{-1}$. 

\item[(A4)] $\left\{\left(X^*_{i j}, i=1, \cdots, n\right): 1 \leq j \leq m\right\}$ is $\alpha$-mixing with $\alpha_{X^*}(s) \leq C \delta^s$, where $\delta \in(0,1)$ and $C>0$ is some constant.
\end{itemize}

Assumptions (A1)--(A3) coincide with the conditions in Theorem~S2 of \citet{feng2024asymptotic}, which essentially require that the dependence among coordinates is not excessively strong.
In particular, the Gaussian requirement in (A1) is mainly imposed to exploit conditional independence properties specific to Gaussian vectors when establishing the joint limit theory.
Relaxing this Gaussianity assumption to accommodate general non-Gaussian designs is an interesting and challenging direction for future work.
Nevertheless, our simulation results indicate that the proposed procedures remain accurate and powerful under a range of non-Gaussian distributions.

Assumption (A4) matches condition (A3) in \citet{ma2024adaptive} and is introduced to facilitate the use of expected shortfall (ES) limit theory for deriving the asymptotic normality of the relevant $L$-statistics.

We now state our first main result, which is about the limiting distribution of each $W_{(j)}$ and the joint limiting distribution of all $W_{(j)}$'s.

\begin{theorem}\label{th1}
	Suppose Assumptions (A1)-(A4) hold. Then as $\min(n,p)\to\infty$, we have
	
	\noindent(\romannumeral1) for all integer $1\le s\le m$ and $x\in\mathbb{R}$,
	\begin{align*}
	\Pr_{H_0}\left(W_{(s)}-b_m\le x \right)\to \Lambda(x)\sum_{i=0}^{s-1}\frac{\{\log \Lambda^{-1}(x)\}^i}{i!},
	\end{align*}
	\noindent(\romannumeral2) for all integer $2\le k\le m$ and $x_1\ge \ldots\ge x_k\in\mR$
	\begin{align*}
	&\Pr_{H_0}\left(\bigcap_{j=1}^k\left(W_{(j)}-b_m \leq x_j\right)\right)\\
	\to&\Lambda\left(x_k\right) \sum_{\sum_{i=2}^j k_i \leq j-1, j=2, \ldots, k} \prod_{i=2}^k \frac{\{\log \Lambda^{-1}(x_i)-\log \Lambda^{-1}(x_{i-1})\}^{k_i}}{k_i!},
	\end{align*}
	where $b_m=2\log m-\log (\log m)$ and $\Lambda(x)=\exp\{-\pi^{-1/2}\exp (-x/2)\}$.
\end{theorem}

According to Theorem \ref{th1}, through convolution, for fixed $k$ the asymptotic null distribution of $L_k$ exists. Specially, $\Pr_{H_0}(L_1-b_m\le x)\to \Lambda(x)$. When $k>1$, the asymptotic null distribution of $T_k$ is complex. So we adapt the bootstrap method to get the null distribution of $L_k$.

We generate bootstrap responses via a wild bootstrap scheme:
\begin{align}
Y_i^{*}= \X_{ia}^\top \hat{\bm\beta}_a+\epsilon_i\,\hat{\varepsilon}_i,\qquad i=1,\ldots,n, \label{eq:wildboot}
\end{align}
where $\hat{\bm\beta}_a=(\X_a^\top\X_a)^{-1}\X_a^\top\Y$ is the least-squares estimator under $H_0$,
$\hat{\varepsilon}_i=Y_i-\X_{ia}^\top\hat{\bm\beta}_a$ are the corresponding residuals, and
$\{\epsilon_i\}_{i=1}^n$ are i.i.d.\ Rademacher random variables taking values $\pm1$ with probability $1/2$,
independent of the data.
Given the bootstrap sample $\{(\X_{ia},Y_i^{*})\}_{i=1}^n$, we compute the bootstrap version of the test statistic,
denoted by $L_{k,l}^{*}$ for the $l$-th bootstrap replication.
Repeating \eqref{eq:wildboot} independently $B$ times yields $\{L_{k,l}^{*}\}_{l=1}^B$, and the Monte Carlo
$p$-value for $L_k$ is computed as
\begin{align}\label{pf}
p_k=\frac{1}{B}\sum_{l=1}^{B}\mathbf{1}\!\left(L_{k,l}^{*}\ge L_k\right).
\end{align}

\subsection{Diverging parameter $k=\lceil \gamma m\rceil$}

To detect dense alternatives, we consider a test based on $L_k$ with diverging $k=\lceil \gamma m\rceil$ and fixed $\gamma\in(0,1)$.
Define $\bm U=(U_1,\cdots,U_m)^\top \sim N(\bm 0, \bms_{b|a})$. Let $\hat{v}_\gamma=U^2_{(\lceil(1-\gamma)m\rceil+1)}$ be the $(1-\gamma)$th sample quantile of $\{U_i^2\}_{1\le i\le m}$, and $v_\gamma$ be the $(1-\gamma)$th quantile of $\chi^2_1$. Define $\beta_j^*=\sum_{l=1}^m \sigma_{jl}^*\beta_{l,b}, j=1,\cdots,m$ and the corresponding top-$k$ order statistics is $\beta_{(1)}^*,\cdots,\beta_{(m)}^*$.

\begin{theorem}\label{th2}
	Under Assumptions (A1)-(A4), for all { fixed} $\gamma_l\in(0,1),l=1,\ldots,s$ with fixed $s$, if $p=o(n^2)$ and $\gamma_l\log(p)n\sum_{i=1}^{m}\beta^{*2}_{(i)}=o(1)$, then as $\min(n,p)\to\infty$,
	\begin{align*}
	\left(\frac{L_{\lceil \gamma_1 m\rceil}-\mu_{\gamma_1}-n\sum_{i=1}^{\lceil\gamma_1 m\rceil}\beta^{*2}_{(i)}}{\sqrt{m\sigma_{\gamma_1\gamma_1}}},\ldots,\frac{L_{\lceil \gamma_s m\rceil}-\mu_{\gamma_s}-n\sum_{i=1}^{\lceil\gamma_s m\rceil}\beta^{*2}_{(i)}}{\sqrt{m\sigma_{\gamma_s\gamma_s}}} \right) \cd \mathcal{N}_s(\bm 0,\bm \Xi),
	\end{align*}
where 
\begin{align*}
\mu_\gamma:=&\sum_{i=1}^m\E_{H_0}\left\{(U_i^2-v_\gamma)\mathbb{I}(U_i^2\ge v_\gamma)+\gamma v_\gamma\right\},\nonumber\\
\sigma_{\gamma_t\gamma_l}:=&\cov_{H_0}\left\{\frac{1}{\sqrt{m}}\sum_{i=1}^m(U_i^2-v_{\gamma_t})\mathbb{I}(U_i^2\ge v_{\gamma_t}),\frac{1}{\sqrt{m}}\sum_{i=1}^m(U_i^2-v_{\gamma_l})\mathbb{I}(U_i^2\ge v_{\gamma_l})\right\}\nonumber\\
=&\frac{1}{m}\sum_{i=1}^m\cov_{H_0}\{(U^2_i-v_{\gamma_t})\mathbb{I}(U^2_i\ge v_{\gamma_t}), (U^2_i-v_{\gamma_l})\mathbb{I}(U^2_i\ge v_{\gamma_l}) \}\nonumber\\
&+\frac{2}{m}\sum_{1\le i<j\le m}\cov_{H_0}\{(U^2_i-v_{\gamma_t})\mathbb{I}(U^2_i\ge v_{\gamma_t}), (U^2_j-v_{\gamma_l})\mathbb{I}(U^2_j\ge v_{\gamma_l}) \}=O(1),
\end{align*}
and $\bm\Xi=(\Xi_{tq})_{s\times s}$ satisfies for $1\le t,q\le s$, $\Xi_{tq}=\sigma_{\gamma_t\gamma_l}/\sqrt{\sigma_{\gamma_t\gamma_t}\sigma_{\gamma_l\gamma_l}}$.
\end{theorem}

Although the diverging-$k$ statistic $L_{\lceil \gamma m\rceil}$ admits an asymptotic normal approximation under $H_0$,
its limiting mean and variance involve nuisance quantities (e.g., threshold-dependent moments and long-run covariances),
which makes a direct implementation cumbersome in finite samples. Therefore, we also use the same wild bootstrap
procedure in \eqref{eq:wildboot} to approximate the null distribution of $L_{\lceil \gamma m\rceil}$ and to compute its
$p$-value. Specifically, for each bootstrap replication $l=1,\ldots,B$, we compute the bootstrap counterpart
$L_{\lceil \gamma m\rceil,l}^{*}$ based on the bootstrap sample $\{(\X_{ia},Y_i^{*})\}_{i=1}^n$, and then define
\begin{align}\label{pd}
p_{\lceil \gamma m\rceil}
=\frac{1}{B}\sum_{l=1}^{B}\mathbf{1}\!\left(L_{\lceil \gamma m\rceil,l}^{*}\ge L_{\lceil \gamma m\rceil}\right).
\end{align}
Since the fixed-$k$ statistics are already evaluated via the same bootstrap routine, incorporating
$L_{\lceil \gamma m\rceil}$ for multiple values of $\gamma$ requires little additional coding and yields a unified and
stable calibration across different choices of $k$.

\subsection{Adaptive procedure}

In practice, the sparsity level of the alternative is unknown, and thus the choice of the truncation parameter $k$ in the $L$-statistic is inherently data-dependent.
A small $k$ is typically more sensitive to  {sparse} alternatives (a few strong coordinates), whereas a larger $k$ tends to be more effective under  {dense} alternatives (many weak coordinates).
Since no single $k$ is uniformly optimal across these regimes, it is natural to construct an adaptive procedure by combining evidence from a collection of $L$-statistics $\{L_k\}$ over a range of $k$ values.

However, such a multi-$k$ aggregation requires understanding the dependence structure among $\{L_k\}$  under the null hypothesis.
Without a precise characterization of these joint relationships, naive combination may lead to overly conservative or inflated type-I error due to strong correlations across nearby $k$'s.
The next theorem addresses this issue by establishing the joint weak convergence of the relevant order-statistic process and a standardized $L$-statistic indexed by a fixed $\gamma$, which in turn implies an asymptotic independence that justifies combining multiple $k$'s in a principled manner.

\begin{theorem}\label{th3}
	Suppose Assumptions (A1)--(A4) hold. If $p=o(n^2)$, then as $\min(n,p)\to\infty$, under $H_0$, we have
	\noindent(\romannumeral1) for all integer $1\le s\le p$ and fixed $\gamma\in(0,1)$,
	\begin{align*}
	\Pr_{H_0}\left(W_{(s)}-b_m\le x, \frac{L_{\lceil \gamma m\rceil}-\mu_{\gamma}}{\sqrt{m\sigma_{\gamma\gamma}}}\le y \right)\to \Lambda(x)\sum_{i=0}^{s-1}\frac{\{\log \Lambda^{-1}(x)\}^i}{i!}\Phi(y),
	\end{align*}
	
	\noindent(\romannumeral2) for all integer $2\le k\le p$ and fixed $\gamma\in(0,1)$,
	\begin{align*}
	&\Pr_{H_0}\left(\bigcap_{j=1}^k\left(W_{(j)}-b_m \leq x_j\right),\frac{L_{\lceil \gamma m\rceil}-\mu_{\gamma}}{\sqrt{m\sigma_{\gamma\gamma}}}\le y\right)\\
	&\to \Lambda\left(x_k\right) \sum_{\sum_{i=2}^j k_i \leq j-1, j=2, \ldots, k} \prod_{i=2}^k \frac{\{\log \Lambda^{-1}(x_i)-\log \Lambda^{-1}(x_{i-1})\}^{k_i}}{k_i!}\Phi(y).
	\end{align*}
\end{theorem}

Motivated by Theorem~\ref{th3}, we aggregate information from a collection of $L$-statistics indexed by a dyadic sequence of truncation levels.
We choose
\[
k_i \;=\; \left\lceil 2^{-i} m \right\rceil,\qquad i=1,\ldots,K,
\quad\text{where}\quad
K \;=\; \left\lfloor \frac{\log(m/20)}{\log 2}\right\rfloor,
\]
so that the grid spans a range from moderately sparse to relatively dense regimes while keeping the smallest truncation at least of order $20$.
For each selected $k_i$, let $p_{{k_i}}$ be the corresponding $p$-value for $L_{k_i}$ (computed as described in (\ref{pd})).
In addition, let $p_{1}$ and $p_{10}$ denote the $p$-values associated with two fixed extreme-type components (e.g., $W_{(1)}$ and $W_{(10)}$; see (\ref{pf}) for precise definitions).
We then define the Cauchy combination statistic
\begin{align}\label{Cauchy combine test}
T_C
&=\frac{1}{K+2}\tan\!\left\{\Big(\frac{1}{2}-p_{1}\Big)\pi\right\}
+\frac{1}{K+2}\tan\!\left\{\Big(\frac{1}{2}-p_{10}\Big)\pi\right\}
+\frac{1}{K+2}\sum_{i=1}^K \tan\!\left\{\Big(\frac{1}{2}-p_{{k_i}}\Big)\pi\right\}.
\end{align}

Theorem~\ref{th3} implies that, under $H_0$, the standardized $L$-statistic is asymptotically independent of the extreme-value component generated by the upper order statistics of $\{W_j\}$, and the same decoupling extends to finite collections indexed by fixed $\gamma$. So we reject the null hypothesis if $T_C\le \alpha$. Consequently, combining $\{p_{T_{k_i}}\}_{i=1}^K$ with the two extreme-type $p$-values via the Cauchy transform in \eqref{Cauchy combine test} provides a robust and powerful omnibus procedure across a wide range of alternatives.
Moreover, the Cauchy combination is known to retain good power when only a small subset of component tests is significant, while remaining stable when signals are more spread out, making it well-suited for adaptivity across sparsity levels.

\section{Simulation}

In this section, we compare our method (abbreviated as CC) with the following four methods: (1) the MAX test proposed by \cite{feng2024asymptotic}; (2) the EB test proposed by \cite{GoemanVanDeGeerVanHouwelingen2006JRSSB}; (3) the PF test proposed by \cite{LanWangTsai2014AISM}; (4) the COM test proposed by \cite{feng2024asymptotic}.

We generate the data from (\ref{model}), where the regression coefficients $\beta_j$ for $j \in\{1,2, \cdots, q\}$ are simulated from a standard normal distribution, and then we set $\beta_j=0$ for $j>q$. In addition, the predictor vector is given by $\boldsymbol{X}_i=\boldsymbol{\Sigma}^{1 / 2} \boldsymbol{z}_i$ for $i=1 \cdots, n$, and each component of $\boldsymbol{z}_i$ is independently generated from three distributions: (i) the normal distribution $N(0,1)$; (ii) the exponential distribution $\exp (1)-1$; (iii) the mixture normal distribution $V / \sqrt{1.8}$, where $V$ is as $0.9N(0,1)+0.1N(0,9)$. We report empirical sizes at the nominal level $0.05$ based on $1000$ Monte Carlo replications.
We consider $(n,p)\in\{100,200\}\times\{200,400,600\}$ and fix $q=5$ baseline covariates.
The covariates have an AR(1) correlation structure $\bms=(0.7^{|i-j|})_{1\le i,j\le p}$.

Table \ref{tab1} reports the empirical sizes of the above five tests.
Overall,  {MAX} is conservative across all settings, with empirical sizes well below $5\%$ in most cases.
Both  {EB} and  {PF} show satisfactory size control under the Gaussian design and remain broadly acceptable under the non-Gaussian designs, although mild inflation is observed in some heavy-tailed or skewed configurations.
The  {COM} procedure delivers the most stable type-I error control across the three designs but is moderately conservative, consistent with its more stringent combination rule.
In contrast,  {CC} can control the emprical sizes very well in all cases.

\begin{table}[htbp]
\centering
\caption{Empirical sizes of five test procedures.}\label{tab1}
\begin{tabular}{c*{9}{r}}
\toprule
 & \multicolumn{3}{c}{(i)} & \multicolumn{3}{c}{(ii)}& \multicolumn{3}{c}{(iii)} \\ \hline
 $p$ &200 &400&600 &200 &400&600 &200 &400&600\\
\midrule
\multicolumn{10}{c}{$n=100$}\\ \hline
MAX  & 2.7 & 3.0 & 1.8 & 3.0 & 2.0 & 2.9 & 2.9 & 2.8 & 2.3 \\
EB  & 4.8 & 5.3 & 3.4 & 4.4 & 4.7 & 5.3 & 5.8 & 6.2 & 5.4 \\
PF  & 5.7 & 5.4 & 3.8 & 5.5 & 4.4 & 6.1 & 5.5 & 6.7 & 5.5 \\
COM  & 3.7 & 3.9 & 2.2 & 4.1 & 3.7 & 4.3 & 4.4 & 5.0 & 4.4 \\
CC  & 5.9 & 6.0 & 4.4 & 5.2 & 6.3 & 6.2 & 5.9 & 5.5 & 5.0 \\ \hline
\multicolumn{10}{c}{$n=200$}\\ \hline
MAX  & 3.4 & 3.5 & 2.9 & 2.5 & 2.5 & 1.8 & 3.5 & 2.9 & 2.1 \\
EB  & 4.2 & 4.8 & 4.9 & 6.0 & 4.9 & 4.8 & 5.1 & 5.4 & 3.7 \\
PF  & 4.6 & 4.9 & 4.9 & 6.6 & 5.3 & 4.9 & 5.8 & 5.5 & 3.9 \\
COM  & 3.6 & 4.7 & 4.1 & 4.3 & 4.4 & 3.0 & 4.6 & 4.6 & 3.4 \\
CC & 5.2 & 5.5 & 5.1 & 6.1 & 5.4 & 5.3 & 6.4 & 6.9 & 5.3 \\
\bottomrule
\end{tabular}
\end{table}

We compare the power of the tests with $n=100, p=200$. Define $\boldsymbol{\beta}_b=\kappa \cdot\left(\beta_{q+1}, \cdots, \beta_p\right)^T$. Let $s$ denote the number of nonzero coefficients. For $s=1, \cdots, m$, we consider $\beta_j \sim N(0,1), q< j \leqslant q+s$ and $\beta_j=0, j>q+s$. The parameter $\kappa$ is chosen so that $\left\|\boldsymbol{\beta}_b\right\|^2=0.8$.  

\begin{figure}[htbp]
\centering
\includegraphics[width=0.32\textwidth]{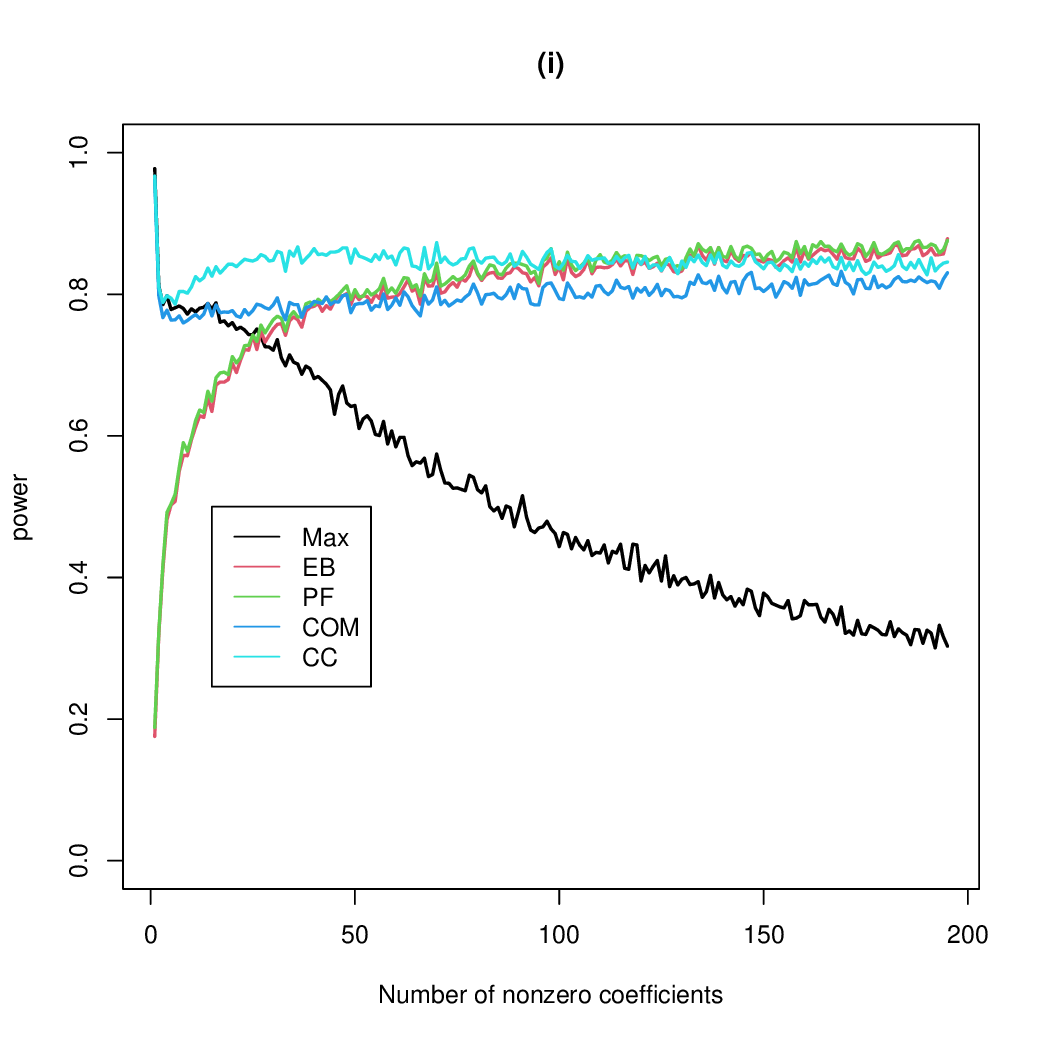}\hfill
\includegraphics[width=0.32\textwidth]{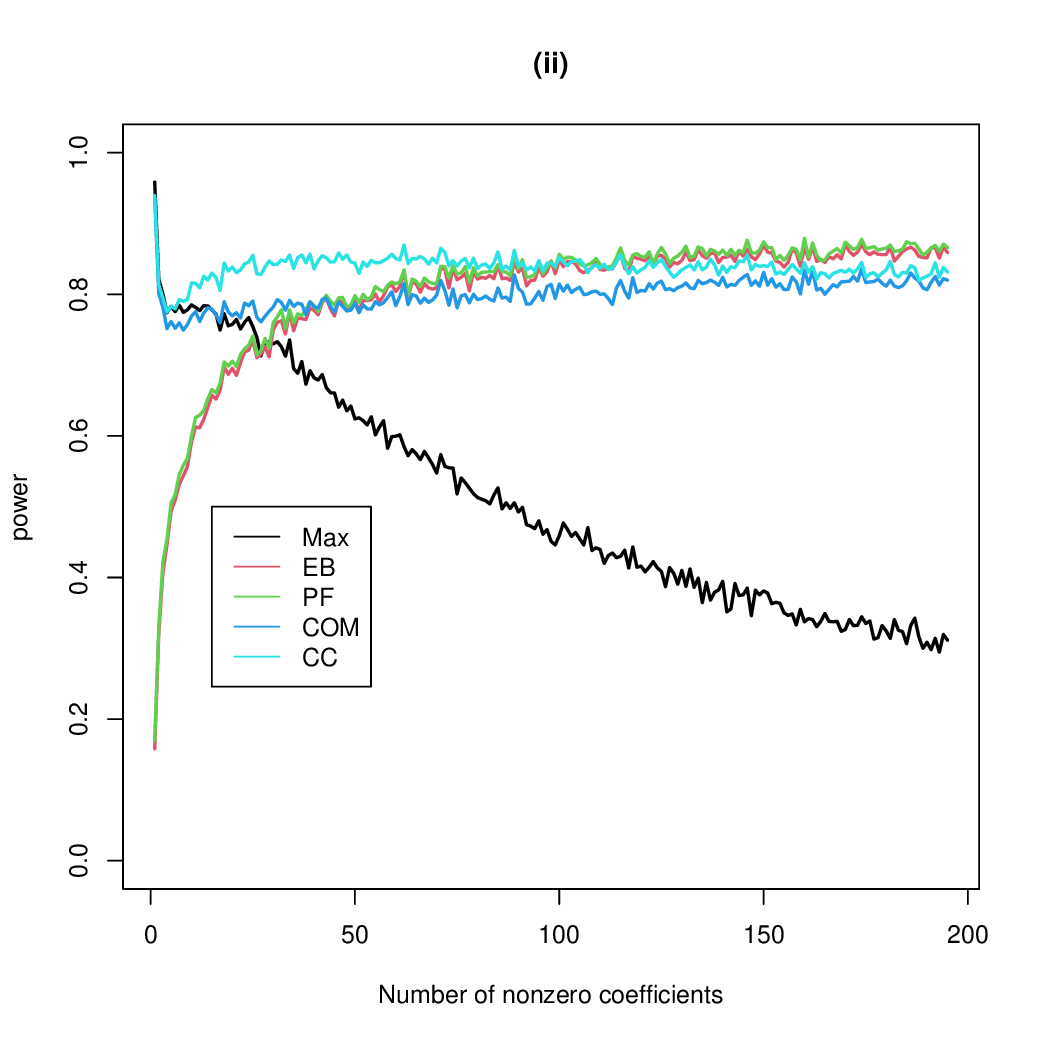}\hfill
\includegraphics[width=0.32\textwidth]{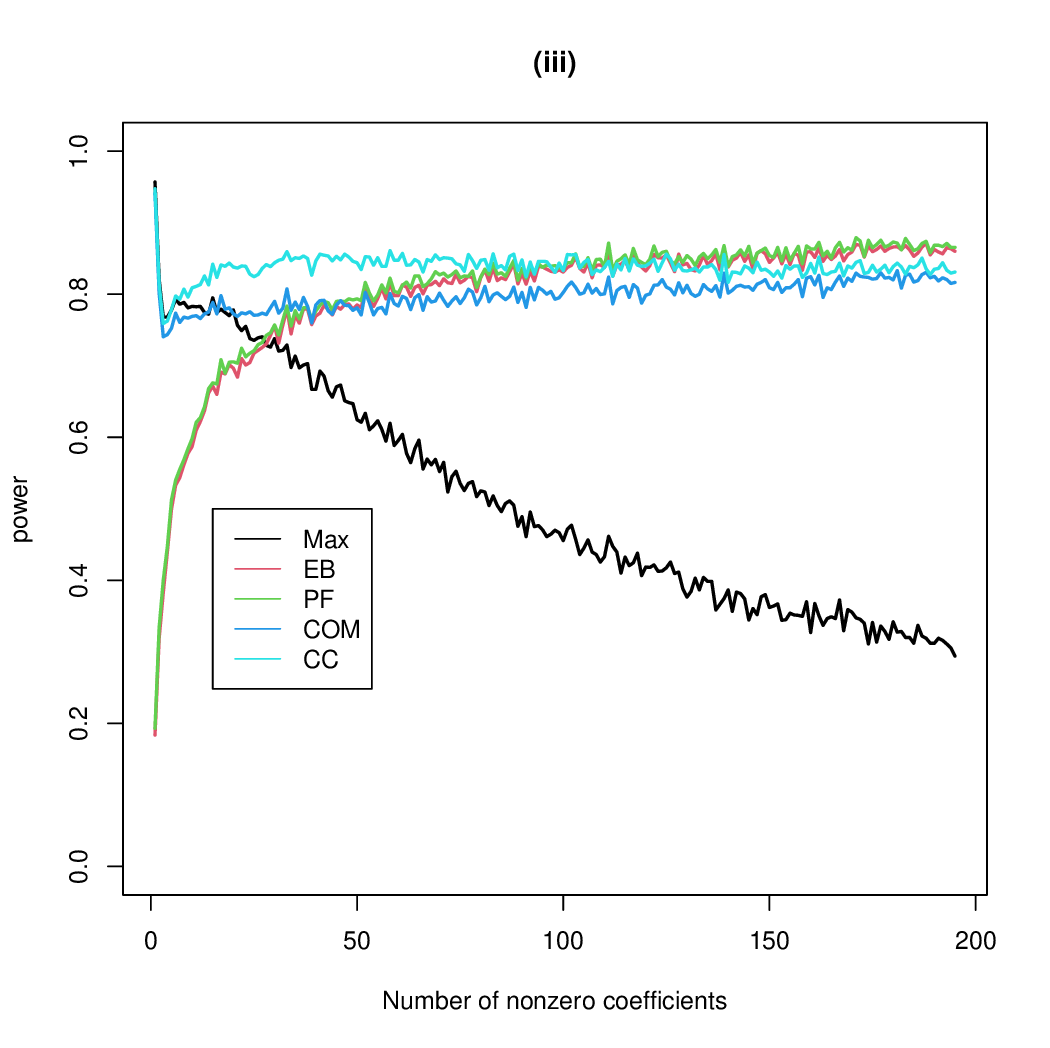}
\caption{Power curves of each test with different distributions under $n=100,p=200$.}
\label{fig1}
\end{figure}

Figure~\ref{fig1} reports the empirical power as a function of the number of nonzero coefficients $s$.
A clear sparsity--density tradeoff is observed.
When the alternative is extremely sparse (very small $s$), the max-type procedure (Max) achieves the highest power, reflecting its sensitivity to a few strong signals.
However, its power decreases steadily as $s$ increases and becomes substantially inferior under moderately sparse to dense alternatives.
In contrast, EB and PF exhibit low power for very small $s$ but increase rapidly with $s$ and then stabilize at a high level once the signal becomes moderately dense, indicating their advantage in aggregating many weak-to-moderate effects.
The combined procedures COM and CC are more robust across sparsity regimes: their power curves remain relatively stable over a wide range of $s$, with CC being uniformly competitive and often the best (or close to the best) for moderate and large $s$, while COM is slightly more conservative but still maintains strong power.
Overall, the results support the adaptivity motivation: max-type tests excel for sparse signals, sum/aggregation-type tests excel for dense signals, and the combination strategy (especially CC) provides consistently strong performance when the sparsity level is unknown.

\section{Conclusion}
In this paper, we develop a family of $L$-statistic procedures for testing regression coefficients in high-dimensional linear models and propose an adaptive omnibus test based on a Cauchy combination.
Both theory and numerical studies indicate that the resulting test attains strong and stable performance across a broad range of sparsity levels under the alternative.

Several directions merit further investigation.
First, part of our limit theory relies on a Gaussianity assumption for the covariates $X$, which is mainly used to simplify dependence arguments.
It would be of substantial interest to relax this requirement and establish analogous results for more general non-Gaussian designs, for instance under elliptical or other weak-dependence frameworks \citep{guo2016robust}.
Second, our current development imposes moment conditions on the regression errors to control the tail behavior of the coordinate-wise evidence measures.
A natural extension is to accommodate substantially heavier-tailed errors by replacing the classical score-type construction with robust alternatives.
In particular, one may incorporate rank-based score tests, such as those proposed by \citet{feng2013rank,xu2017new,xu2021maximum}, to improve robustness while retaining power in high dimensions. Finally, it is of interest to extend our framework beyond the linear model. One natural direction is to develop counterparts for generalized linear models, following the global-testing line of work in \citet{GoemanVanDeGeerVanHouwelingen2006JRSSB} and the more recent advances in \citet{yang2024tests}. Another promising direction is to adapt the proposed $L$-statistic and combination strategy to nonlinear regression settings, including modern testing procedures for nonparametric or kernel-based models; see, for instance, \citet{li2023testing} and \citet{yin2026kernel}.

\section{Appendix}
\subsection{Proof of Theorem \ref{th1}}
Define $\boldsymbol{u}_j=(X^*_{1j},\cdots,X_{nj}^*)^\top$ and $\boldsymbol{e}_2=\left(\mathbf{I}_n-\mathbf{H}_a\right) \boldsymbol{\varepsilon} /\left\|\left(\mathbf{I}_n-\mathbf{H}_a\right) \boldsymbol{\varepsilon}\right\|$. By assumption (A1), $\left\{\mathbf{u}_j ; 1 \le j \le m\right\}$ are independent of $\boldsymbol{X}_a$ and $\boldsymbol{\varepsilon}$, and hence are independent of the unit random vector $\mathbf{e}_2$. We then have from Lemma S16 in \cite{feng2024asymptotic} that $\left(\mathbf{u}_{1}^T \mathbf{e}_2, \cdots, \mathbf{u}_m^T \mathbf{e}_2\right)^T$ has distribution $N\left(\mathbf{0}, \boldsymbol{\Sigma}_{b | a}\right)$ and is also independent of $\left\|\left(\mathbf{I}_n-\mathbf{H}_a\right) \boldsymbol{\varepsilon}\right\|$.  Define $\tilde W_{(k)}$ is the top-$k$ order statistic of $\{\tilde Z_i^2\}_{i=1}^m$ and $\tilde Z_i=\u_i^\top \e_2$. 

Following the same decomposition as in the proof of Theorem S2 of \citet{feng2024asymptotic},
one may represent $\tilde \X_{\cdot j}=(\I_n-\H_a)\u_j$ with $\u_j\sim N(0,\I_n)$.
Then the key identity
\begin{equation}\label{eq:S164_like}
(\tilde \X_{\cdot j}^\top {\bm\varepsilon})^2
=\|(\I_n-\H_a){\bm \varepsilon}\|^2\cdot (\u_j^\top \e_2)^2
\end{equation}
holds. Using $\hat{\bm \varepsilon}=(\I_n-\H_a){\bm\varepsilon}$ and $\hat\sigma^2=r^{-1}\|(\I_n-\H_a){\bm\varepsilon}\|^2$, $r=n-q$,
we get
\[
Z_j^2
=\frac{(\tilde \X_{\cdot j}^\top \hat{\bm \varepsilon})^2}{\hat\sigma^2\,\|\tilde \X_{\cdot j}\|^2}
=\frac{\|(\I_n-\H_a)\bm \varepsilon\|^2 (\u_j^\top \e_2)^2}{\hat\sigma^2\,\|\tilde \X_{\cdot j}\|^2}
=\frac{r\,(\u_j^\top \e_2)^2}{\|\tilde \X_{\cdot j}\|^2}.
\]
Define
$A_j:= r^{-1}\|\tilde \X_{\cdot j}\|^2.$
Then we obtain the convenient representation
\begin{equation}\label{eq:Wj_Aj}
W_j = Z_j^2 = \frac{\tilde W_j}{A_j}.
\end{equation}
Conditional on $X_a$ (equivalently, on $\P=\I_n-\H_a$), $\P \u_j$ is Gaussian with covariance $\P$.
Since $\P$ is an orthogonal projection of rank $r$, we have the standard identity
\begin{equation}\label{eq:chi_square}
rA_j=\|P u_j\|^2 \ \stackrel{d}{=} \ \chi_r^2,
\qquad \text{conditionally on }\P.
\end{equation}
We next show that $\max_{1\le j\le m}|A_j-1|$ is small.
By the Laurent--Massart tail bound for $\chi_r^2$, for any $t>0$,
\[
\Pr\!\left(\chi_r^2-r\ge 2\sqrt{rt}+2t\right)\le e^{-t},
\qquad
\Pr\!\left(r-\chi_r^2\ge 2\sqrt{rt}\right)\le e^{-t}.
\]
Applying this to \eqref{eq:chi_square} gives, for each fixed $j$,
\[
\Pr\!\left(|A_j-1|\ge 2\sqrt{\frac{t}{r}}+\frac{2t}{r}\right)\le 2e^{-t}.
\]
Let $t=c\log m$ with a constant $c>2$.
By a union bound over $j=1,\ldots,m$,
\begin{equation}\label{eq:uniform_Aj}
\Pr\!\left(\max_{1\le j\le m}|A_j-1|\ge \delta_{m,r}\right)
\le 2m e^{-c\log m}=2m^{\,1-c}\to 0,
\end{equation}
where
\begin{equation}\label{eq:delta_mr}
\delta_{m,r}:=2\sqrt{\frac{c\log m}{r}}+\frac{2c\log m}{r}.
\end{equation}
Hence,
\begin{equation}\label{eq:delta_op}
\Delta_n:=\max_{1\le j\le m}|A_j-1|=O_p\!\left(\sqrt{\frac{\log m}{r}}\right).
\end{equation}
On the event $\{\Delta_n\le 1/2\}$, for every $j$ we have
$1-\Delta_n\le A_j\le 1+\Delta_n$, hence
\[
\frac{1}{1+\Delta_n}\tilde W_j
\le \frac{\tilde W_j}{A_j}
\le \frac{1}{1-\Delta_n}\tilde W_j.
\]
Using \eqref{eq:Wj_Aj} and the monotonicity of order statistics,
this componentwise inequality implies the  {same} inequality for $k$-th order statistics:
\begin{equation}\label{eq:order_sandwich}
\frac{1}{1+\Delta_n}\tilde W_{(k)}
\le W_{(k)}
\le \frac{1}{1-\Delta_n}\tilde W_{(k)}.
\end{equation}
Therefore,
\begin{equation}\label{eq:diff_bound}
|W_{(k)}-\tilde W_{(k)}|
\le \max\!\left\{\frac{1}{1-\Delta_n}-1,\ 1-\frac{1}{1+\Delta_n}\right\}\tilde W_{(k)}
\le \frac{\Delta_n}{1-\Delta_n}\tilde W_{(k)}
\le 2\Delta_n\,\tilde W_{(k)},
\end{equation}
where the last inequality holds on $\{\Delta_n\le 1/2\}$.
We show $\tilde W_{(k)}=O_p(\log m)$, which is enough for fixed $k$.
Indeed, by a union bound (no independence is needed),
for any $t>0$,
\[
\Pr\!\left(\tilde W_{(1)}>t\right)
=\Pr\!\left(\max_{1\le j\le m}\tilde Z_j^2>t\right)
\le \sum_{j=1}^m \Pr(\tilde Z_j^2>t)
\le m\Pr(\chi_1^2>t).
\]
Using $\Pr(\chi_1^2>t)\le 2e^{-t/2}$ for large $t$, we obtain
$\Pr(\tilde W_{(1)}>C\log m)\le 2m^{1-C/2}\to 0$ as soon as $C>2$.
Hence $\tilde W_{(1)}=O_p(\log m)$, and therefore
\begin{equation}\label{eq:tildeWk_logm}
\tilde W_{(k)}\le \tilde W_{(1)}=O_p(\log m).
\end{equation}
Combining \eqref{eq:diff_bound}, \eqref{eq:delta_op}, and \eqref{eq:tildeWk_logm} yields
\[
|W_{(k)}-\tilde W_{(k)}|
=O_p\!\left(\sqrt{\frac{\log m}{r}}\right)\cdot O_p(\log m)
=O_p\!\left(\frac{(\log m)^{3/2}}{\sqrt r}\right).
\]
Since $r=n-q\asymp n$ and $m=o(n^2)$, we have $\log m=O(\log n)$ and thus
$(\log m)^{3/2}/\sqrt r\to 0$, implying 
\begin{align}\label{wk}
|W_{(k)}-\tilde W_{(k)}|=o_p(1).
\end{align}
Consequently, under Assumptions (A1)--(A4), the conditions required by Theorem~1 of \citet{ma2024adaptive} are satisfied, and the desired limit result follows immediately by applying that theorem. \hfill$\Box$

\subsection{Proof of Theorem \ref{th2}}

Conditional on $\H_a$, $\tilde \X_{\cdot j}=(\I_n-\H_a)\u_j$ is Gaussian with covariance $(\I_n-\H_a)$,
a rank-$r$ orthogonal projection. Hence
\[
rA_j = \|\tilde \X_{\cdot j}\|^2 \ \stackrel{d}{=}\ \chi_r^2,
\qquad \text{conditionally on }\H_a.
\]
Therefore,
\begin{equation}\label{eq:Aj_moments}
E(A_j\mid \H_a)=1,\qquad \var(A_j\mid \H_a)=\frac{2}{r},
\end{equation}
and  for $r$ large enough,
\begin{equation}\label{eq:Aj_inv_moments}
E\!\left[(A_j^{-1}-1)^2\mid \H_a\right]=O(r^{-1}),\qquad
E\!\left[|A_j^{-1}-1|\mid \H_a\right]=O(r^{-1}).
\end{equation}
In addition, the proof of Theorem S2 in \cite{feng2024asymptotic} establishes a uniform control on these norm terms,
namely $\max_{q+1\le j\le p}\big|1-(r^{-1}\|\tilde \X_{\cdot j}\|^2)^{-1}\big|$ is negligible
at a logarithmic rate, which implies that multiplying/dividing by $A_j$
does not change the order-statistic structure in any essential way.

For fixed $\gamma\in(0,1)$, the top-$\lceil\gamma m\rceil$ sum is an expected-shortfall functional.
Assumption (A4) is imposed precisely to invoke an ES limit theory for $\{W_j\}$ and $\{\tilde W_j\}$.
In particular, one can use the ES approximation step used for deriving the asymptotic normality of
$L_{\lceil\gamma m\rceil}$ \citep{ma2024adaptive},
to write
\begin{equation}\label{eq:ES_representation}
L_{\lceil\gamma m\rceil}
=\sum_{j=1}^m W_j\,\mathbf 1\!\left(W_j\ge \xi_\gamma\right)+o_p(m^{1/2}),
\qquad
\tilde L_{\lceil\gamma m\rceil}
=\sum_{j=1}^m \tilde W_j\,\mathbf 1\!\left(\tilde W_j\ge \tilde \xi_\gamma\right)+o_p(m^{1/2}),
\end{equation}
where $\xi_\gamma,\tilde\xi_\gamma$ are the corresponding population $(1-\gamma)$-quantiles.

By \eqref{eq:Wj_Aj},
\[
W_j-\tilde W_j = \tilde W_j(A_j^{-1}-1).
\]
Plugging this into \eqref{eq:ES_representation} and using the uniform smallness of $A_j-1$,
we reduce the problem to showing
\begin{equation}\label{eq:core_sum}
\sum_{j=1}^m \tilde W_j(A_j^{-1}-1)\mathbf 1(\tilde W_j\ge \tilde\xi_\gamma)
=o_p(m^{1/2}).
\end{equation}
Let
\[
\Delta_j := \tilde W_j(A_j^{-1}-1)\mathbf 1(\tilde W_j\ge \tilde\xi_\gamma).
\]
By Cauchy--Schwarz and \eqref{eq:Aj_inv_moments},
\[
E(\Delta_j^2\mid H_a,e_2)
\le E(\tilde W_j^2\mid e_2)\cdot E((A_j^{-1}-1)^2\mid H_a)
= O(1)\cdot O(r^{-1})=O(r^{-1}),
\]
since $\tilde W_j=\tilde Z_j^2$ has bounded second moments uniformly.
Also $\E(\Delta_j\mid H_a,e_2)=O(r^{-1})$ by \eqref{eq:Aj_inv_moments}.
Under (A1)--(A3), the dependence across $j$ is weak enough so that
\[
\var\!\left(\sum_{j=1}^m \Delta_j \,\middle|\, H_a,e_2\right)
= O\!\left(\frac{m}{r}\right),
\qquad
E\!\left(\sum_{j=1}^m \Delta_j \,\middle|\, H_a,e_2\right)
= O\!\left(\frac{m}{r}\right).
\]
Hence, by Chebyshev's inequality,
\[
\sum_{j=1}^m \Delta_j
= O_p\!\left(\frac{m}{r}\right)+O_p\!\left(\sqrt{\frac{m}{r}}\right).
\]
Since $r=n-q\asymp n$ and $m=o(n^2)$, we have
\[
\frac{m/r}{m^{1/2}}=\frac{m^{1/2}}{r}=o(1),
\qquad
\frac{\sqrt{m/r}}{m^{1/2}}=\frac{1}{\sqrt r}=o(1),
\]
which proves \eqref{eq:core_sum}. Combining this with \eqref{eq:ES_representation}
yields 
\begin{align}\label{wg}
L_{\lceil \gamma m\rceil}=\tilde{L}_{\lceil \gamma m\rceil}+o_p(m^{1/2}),
\end{align}
where $\tilde{L}_{\lceil \gamma m\rceil}=\sum_{j=1}^{{\lceil \gamma m\rceil}} \tilde{W}_{(j)}$. Then, because $(\tilde Z_1,\cdots,\tilde Z_m)^\top \sim N(\bm 0, \bms_{b|a})$, we then obtain the result by Theorem 3 in \cite{ma2024adaptive}. \hfill$\Box$

\subsection{Proof of Theorem \ref{th3}}
By (\ref{wk}) and (\ref{wg}), we can easily obtain the result by Theorem 4 in \cite{ma2024adaptive}. \hfill$\Box$

\bibliographystyle{chicago}
\bibliography{ref}

@article{feng2024asymptotic,
    author  = {Feng, Long and Jiang, Tiefeng and Li, Xiaoyun and Liu, Binghui},
    title   = {Asymptotic independence of the sum and maximum of dependent random variables with applications to high-dimensional tests},
    journal = {Statistica Sinica},
    year    = {2024},
    volume  = {34},
    pages   = {1745--1763},
}

@article{ma2024adaptive,
  title={Adaptive L-statistics for high dimensional test problem},
  author={Ma, Huifang and Feng, Long and Wang, Zhaojun},
  journal={arXiv preprint arXiv:2410.14308},
  year={2024}
}

@article{GoemanVanDeGeerVanHouwelingen2006JRSSB,
  author  = {Goeman, J. J. and van de Geer, S. A. and van Houwelingen, H. C.},
  title   = {Testing against a high dimensional alternative},
  journal = {Journal of the Royal Statistical Society: Series B (Statistical Methodology)},
  year    = {2006},
  volume  = {68},
  number  = {3},
  pages   = {477--493}
}

@article{GoemanVanHouwelingenFinos2011Biometrika,
  author  = {Goeman, J. J. and van Houwelingen, H. C. and Finos, L.},
  title   = {Testing against a high-dimensional alternative in the generalized linear model: asymptotic type {I} error control},
  journal = {Biometrika},
  year    = {2011},
  volume = {98},
  number = {2},
  pages   = {381--390}
}

@article{LanWangTsai2014AISM,
  author  = {Lan, Wenhua and Wang, Hao and Tsai, Chih-Ling},
  title   = {Testing covariates in high-dimensional regression},
  journal = {Annals of the Institute of Statistical Mathematics},
  year    = {2014},
  volume  = {66},
  number  = {2},
  pages   = {279--301}
}

@article{ZhongChen2011JASA,
  author  = {Zhong, P.-S. and Chen, S. X.},
  title   = {Tests for high-dimensional regression coefficients with factorial designs},
  journal = {Journal of the American Statistical Association},
  year    = {2011},
  volume  = {106},
  number  = {493},
  pages   = {260--274}
}

@article{feng2013rank,
  title={Rank-based score tests for high-dimensional regression coefficients},
  author={Feng, Long and Zou, Changliang and Wang, Zhaojun and Chen, Bin},
  journal={Electronic Journal of Statistics},
  volume={7},
  pages={2131--2149},
  year={2013}
}

@article{yang2024tests,
  title={Tests for high-dimensional generalized linear models under general covariance structure},
  author={Yang, Weichao and Guo, Xu and Zhu, Lixing},
  journal={Computational Statistics \& Data Analysis},
  volume={199},
  pages={108026},
  year={2024},
  publisher={Elsevier}
}

@article{li2023testing,
  title={Testing the effects of high-dimensional covariates via aggregating cumulative covariances},
  author={Li, Runze and Xu, Kai and Zhou, Yeqing and Zhu, Liping},
  journal={Journal of the American Statistical Association},
  volume={118},
  number={543},
  pages={2184--2194},
  year={2023},
  publisher={Taylor \& Francis}
}

@article{xu2021maximum,
  title={Maximum-type tests for high-dimensional regression coefficients using Wilcoxon scores},
  author={Xu, Kai and Zhou, Yeqing},
  journal={Journal of Statistical Planning and Inference},
  volume={211},
  pages={221--240},
  year={2021},
  publisher={Elsevier}
}

@article{xu2017new,
  title={A new nonparametric test for high-dimensional regression coefficients},
  author={Xu, Kai},
  journal={Journal of Statistical Computation and Simulation},
  volume={87},
  number={5},
  pages={855--869},
  year={2017},
  publisher={Taylor \& Francis}
}

@article{yin2026kernel,
  title={Kernel-based marginal testing for covariate effects in high-dimensional settings},
  author={Yin, Hong and Wang, Yijun and Xu, Ancha},
  journal={Scandinavian Journal of Statistics},
  year={2026},
  publisher={Wiley Online Library}
}

@article{guo2022conditional,
  title={Conditional test for ultrahigh dimensional linear regression coefficients},
  author={Guo, Wenwen and Zhong, Wei and Duan, Sunpeng and Cui, Hengjian},
  journal={Statistica Sinica},
  volume={32},
  number={3},
  pages={1381--1409},
  year={2022},
  publisher={JSTOR}
}

@article{cui2018test,
  title={TEST FOR HIGH-DIMENSIONAL REGRESSION COEFFICIENTS USING REFITTED CROSS-VALIDATION VARIANCE ESTIMATION},
  author={Cui, Hengjian and Guo, Wenwen and Zhong, Wei},
  journal={The Annals of Statistics},
  volume={46},
  number={3},
  pages={958--988},
  year={2018},
  publisher={JSTOR}
}

@article{guo2016robust,
  title={Robust U-type test for high dimensional regression coefficients using refitted cross-validation variance estimation},
  author={Guo, WenWen and Chen, YongShuai and Cui, HengJian},
  journal={Science China Mathematics},
  volume={59},
  number={12},
  pages={2319--2334},
  year={2016},
  publisher={Springer}
}
\end{document}